# Electronic structure of the Ge/Si(105) hetero-interface

P. M. Sheverdyaeva[1], C. Hogan[2], A. Sgarlata[3], L. Fazi[3], M. Fanfoni[3], L. Persichetti[4], P. Moras[1] and A. Balzarotti[3]

[1]*Istituto di Struttura della Materia-CNR (ISM-CNR) S.S. 14, km 163.5, I-34149, Trieste, Italy*
[2]*Istituto di Struttura della Materia-CNR (ISM-CNR), Via Fosso del Cavaliere 100, 00133 Roma, Italy*
[3]*Dipartimento di Fisica, Università di Roma "Tor Vergata", Roma, Italy*
[4]*Dipartimento di Scienze, Università Roma Tre, Viale G. Marconi, 446- 00146 Roma, Italy*

Thin Ge layers deposited on Si(105) form a stable single-domain film structure with large terraces and rebonded-step surface termination, thus realizing an extended and ordered Ge/Si planar hetero-junction. At the coverage of four Ge monolayers angle-resolved photoemission spectroscopy reveals the presence of two-dimensional surface and film bands displaying energy-momentum dispersion compatible with the 5×4 periodicity of the system. The good agreement between experiment and first-principles electronic structure calculations confirms the validity of the rebonded-step structural model. The direct observation of surface features within 1 eV below the valence band maximum corroborates previously reported analysis of the electronic and optical behavior of the Ge/Si hetero-interface.

# I. Introduction

The observation that a germanium film on Si(105) remains flat up to a three times larger thickness than on Si(001) [1], prompted a number of studies aiming to understand the mechanism beneath this behavior. It was recognized that the Ge-covered Si(105) surface (Ge/Si(105)) is rebonded-step (RS)-reconstructed [2, 3] and much more strained than Ge-covered Si(001). Ge relieves the tensile stress of the substrate, nullifying it after about 11 atomic layers of coverage [4] compared to 3.5 atomic layers of Ge/Si(001), in good quantitative agreement with experiments. The electronic and elastic contribution to the surface energy of dimers that form the characteristic horseshoe motif within the reconstructed unit cell (see Fig. 1(f)) was theoretically clarified in considerable detail [5], providing an explanation of the structural stability of the (105) vicinal surface. Experimental work on the Ge/Si(105) epitaxy elucidated the role of the RS reconstruction in determining the structural evolution of strained three-dimensional islands which grow above the critical thickness [6-10].

Recent analysis of the optical properties of Ge/Si(105) by reflectance anisotropy spectroscopy (RAS) provided insight about subsurface bonds and surface band structure [11]. On the basis of first-principles electronic structure calculations, it was found that the observed optical transitions involve surface resonances located well below the valence band maximum (VBM) and unoccupied states localized inside the projected bulk band gap of silicon. These latter states are recognized as surface features with little dispersion along orthogonal symmetry directions of the 5×4 surface Brillouin zone (SBZ). However, an experimental investigation aimed at exploring directly the electronic structure of the Ge/Si(105) interface is lacking to date.

In this paper we report on an angle-resolved photoemission spectroscopy (ARPES) study of Ge/Si(105). We find a number of surface and film states repeating in the reciprocal space according to the periodicity of the 5×4 SBZ. First-principles electronic structure calculations using the rebonded-step structural model reproduce well the experimental observations. This agreement reinforces the interpretation of the RAS data reported in Ref. [11]. We find that some surface features display elongated photoemission patterns on constant energy planes, which are aligned with the $[0\bar{1}0]$ axis. These observations can be linked to the anisotropic electronic coupling among specific Ge atoms of the RS surface reconstruction.

## II. Methods

The Si(105) surface is used for the growth of epitaxial Ge films displaying the RS reconstruction. The Si substrate (n-type, P-doped with resistivity < 0.005 Ω cm) has a miscut angle θ = 11.3° with respect to the (001) plane toward the [100] direction [12]. It was cleaned by flash annealing to 1370 K in ultra-high vacuum conditions. The absence of contaminants was checked by core level measurements. Ge was deposited at room temperature under a constant flux of 0.3 ML/min [1 monolayer (ML) corresponds to $6.3 \times 10^{14}$ atoms/cm$^2$]. The system was annealed to 870 K for 3 minutes to obtain a fully-reconstructed surface and suppress the formation of amorphous Ge clusters. The RS surface showed a clear 5×4 low-energy electron diffraction (LEED) pattern (Fig. 1(c)), corresponding to the RS reconstruction.

ARPES experiments were performed at the VUV-Photoemission beamline of the synchrotron radiation facility Elettra in Trieste (Italy). Photoelectrons excited by linear p-polarized light (45° incidence angle) were recorded by a Scienta R4000-WAL hemispherical electron analyzer at room temperature. Constant energy cuts were acquired by tilting the normal to the sample in the direction perpendicular to the slit of the analyzer. The direction of the miscut was oriented approximately at 45° with respect to the slit of the analyzer. Energy and angular resolutions were set to 25 meV and 0.3°, respectively. The energy scale was referred to the VBM of the Si substrate.

First-principles calculations were performed using density functional theory (DFT) in the local density approximation (LDA) [13-15]. A plane wave basis set was used, with kinetic energy cut-off of 30 Ry, and norm-conserving, scalar relativistic pseudopotentials [16]. A test calculation was also carried out with fully relativistic pseudopotentials to examine the effect of spin-orbit coupling. Nonlinear core corrections were used for Ge [17]. Geometry relaxations were performed using quantum-ESPRESSO [18]. The Ge/Si(105) surface was modeled using supercells containing slabs, 22 Å thick, separated by vacuum regions about 14 Å thick. Dangling bonds at the bottom of the slab were terminated with hydrogen, and several back layers of Si were kept fixed at the bulk positions during structural relaxation. The remaining atoms were relaxed until individual force components were less than 12 meVÅ$^{-1}$. A (2×2×1) *k*-point mesh was used for SBZ sampling. The theoretical lattice constant of bulk silicon, a = 5.431 Å, was adopted in order to replicate the conditions of strain felt by the Ge overlayer. The obtained structure and surface geometry are illustrated in Fig. 1(f,g). Although the surface is vicinal no sharp steps remain after relaxation to the RS reconstruction geometry.

Electronic structure calculations were then performed for the optimized geometry around a closed path within the SBZ of the 5×4 reconstruction. The resulting band structure plotted in the repeated zone scheme is highly folded, mixes states of Ge and Si, and is therefore difficult to analyze. In order to obtain a deeper understanding we computed the density of states (DOS) projected onto Ge and Si atoms, respectively, first integrated over the whole slab $\rho(E)$ using a fine (16×16×1) $k$-point mesh, but also resolved in $k$-space $\rho(E,k_{//})$. The color scale of the $k$-resolved DOS indicates the spatial localization of the states and allows to discriminate between Ge (film and surface) and Si states. We also performed an unfolding process along high-symmetry directions to obtain an effective band structure of the primitive 1×1 cell, using the BandUP software [19]. This allows us to clearly distinguish localized states from dispersive states having ×4, ×5, or ×1 periodicities and is essential for interpreting such highly folded band structures.

### III. Results and Discussion

Fig. 1 describes the structural features of the Si(105) and Ge/Si(105) systems. Fig. 1(a) displays a scanning tunneling microscopy (STM) image of the clean Si(105) surface. Due to the large polar and azimuthal misorientation of the Si(105) plane from the singular Si (001) facet, the formation of a stable 2×1 reconstruction is hindered on the Si(105) face which is, therefore, atomically rough and lacks long-range order [20].

Instead, the 4 ML Ge/Si(105) system [Fig. 1(b)] shows a flat morphology with the single-domain RS superstructure extending over large terraces. In line with previous results [5, 12], these observations indicate that the RS-reconstructed Ge/Si(105) surface becomes a flat singular orientation of the lattice-mismatched heteroepitaxial Ge/Si system which is stabilized by the compressively strained environment. The appearance of long-range order on the surface layers after Ge deposition is confirmed by the experimental LEED pattern, shown in Fig. 1(c), together with the simulated diffraction pattern calculated using three reconstructed atomic layers and the 5×4 SBZ (insets).

The RS surface contains a very large single-reconstruction motif [2, 21] which gives a non-trivial electronic contrast in bias-dependent STM images (Fig. 1(d, e)). The 5×4 unit cell and the characteristic horseshoe motif are shown in Fig. 1(f, g), which represents top and side views of the model Ge/Si(105) interface according to Ref. [22]. The thickness of the 4 ML Ge film is 5.6 Å [23].

Fig. 2(a-c) shows ARPES data for the clean Si(105) surface at 130 eV photon energy. The constant energy cut of Fig. 2(a) (-1.00 eV) is characterized by replicas of a simple pattern, formed by one star-like and two square-like concentric contours (white continuous lines). It is straightforward to associate these contours to the heavy-hole (HH), light-hole (LH) and split-off (SO) bulk valence bands of Si, respectively [24]. The three main replicas of this pattern originate from adjacent bulk Brillouin zones of Si. The surface projection of their centers defines the position of $\bar{\Gamma}$ points, which will be used as a reference for the Ge-related electronic states in the reciprocal space. At smaller energies (-0.13 eV in Fig. 2(b)) the contours shrink, as dictated by the energy-momentum relation of the HH, LH and SO bands (Fig. 2(c)).

Upon formation of the 4 ML Ge film, new electronic states appear in the ARPES data (Fig. 2(d-f)), in line with the high structural quality of the system observed in the STM image of Fig. 1(b). For instance, well-defined structures, elongated in the vertical and horizontal directions, are seen in the region between the principal Si-related contours in Fig. 2(d,e). These features derive from a number of bands indicated by arrows in Fig. 2(f), which are absent in the corresponding region of the disordered Si(105) surface (Fig. 2(c)). The shape of the constant energy contours of Si is modified by the presence of overlapping Ge states, especially at low binding energies. This effect becomes evident when the insets of Fig. 2(b,e) are compared. New star-like (HH band) and square-like (LH band) contours, different from those of clean Si, are clearly observed in the spectra of the Ge/Si(105) system.

In general, the dispersion of the Ge states in Fig. 2(d-f) does not display a simple relation with the repeated 5×4 SBZ (rectangles in Fig. 2(e)), probably because of the limited surface sensitivity achieved with of 130 eV photons. This connection becomes clearer at lower photon energies. Fig. 3(a-d) reports ARPES data for the Ge/Si(105) system measured with 40 eV photons. Fig. 3(a) show a constant energy cut at -0.25 eV, along with symmetry points and axes of the system. Fig. 3(b) shows the second derivative [along the energy axis] of the same data. Fig. 3(c,d) display ARPES scans along selected $\bar{\Gamma}$-$\bar{K}'$ lines of the repeated 5×4 SBZ (indicated by corresponding labels in Fig. 3(a)). These scans are presented in the second derivative form (along the energy axis) to enhance the signal from weak electronic states. Several features are found to replicate in the reciprocal space according to the ×5 periodicity of the SBZ. The topmost part of Fig. 3(c) displays two sine-like bands with opposite phases and band widths of about 200 meV (white dotted lines). The same wavy bands can be recognized along the equivalent direction shown in Fig. 3(d). These bands derive from the folding of the HH and LH Ge-related bands already detected in Fig. 2(e,f), which are very close in energy along the scanned direction.

Confirmation of these analyses is provided by *ab-initio* calculations. Fig. 4(a,b) report the computed band structures of Ge/Si(105) along the $k_y$ direction scanned in Fig. 2(f), before and after the unfolding process. While panel (a) shows a dense tapestry of overlapping bands, the unfolded bands of panel (b) clearly reveal the strongly dispersive HH band of bulk Si (band width ~ 3eV). With respect to bulk Si, there are two main differences. First, the Si HH band is strongly perturbed above 1eV by the presence of the Ge layer such that its character is indeed lost as it approaches the $\overline{\Gamma}$ points. Second, a rich texture of filled bands is visible in the center of the figure near the VBM, consistent with the experimental observation in Fig. 2(f).

In order to isolate surface states from other Ge- and Si-related states, we plot the band structure [reduced zone scheme] of the Ge/Si(105) system in Fig. 4(c) and superimpose those of bulk Si projected onto the 5×4 SBZ. Two occupied "true" surface states are thus identified within the bulk gap along $\overline{K}'$-$\overline{J}$-$\overline{K}$. The higher energy state has a band width of about 150 meV and is split at the $\overline{K}$ point. The other state lying close to $\overline{J}$ instead quickly becomes resonant with the bulk. Inclusion of spin-orbit coupling [see inset] leads only to a slight (< 30 meV) broadening of the bands. Both states contribute to the total and Ge-projected integrated DOS $\rho(E)$ in Fig 4(d), giving rise to two peaks V1 at -0.15 eV and V2 at -0.35 eV, respectively. The Ge-projected DOS also reveals the presence of a strong feature V3 at -0.6 eV and a weaker signal V4 at -0.8 eV, both resonant with the bulk. These four states are resolved in Fig. 4(e). This $\rho(E,k_{\parallel})$ plot reveals that the V2 peak actually derives from states along $\overline{\Gamma}$-$\overline{K}$ having a strong Ge component, and that V3 has a complicated character: a broad, flattish profile throughout the SBZ as well as an overlapping dispersive component (V3') along $\overline{\Gamma}$-$\overline{K}'$, which overlaps in energy with V4.

The topmost bands observed in the ARPES spectra in Fig. 3(c,d), discussed previously, are now shown in Fig. 3(e) as dotted lines overlaid onto the computed $\rho(E,k_{\parallel})$ resolved along $\overline{\Gamma}$-$\overline{K}'$ and repeated along [$\overline{5}$01]. Noting the weak spatial localization on surface Ge atoms, we confirm that the signal derives from the HH/LH band of Si modified strongly by the Ge film and folded according to the ×5 periodicity. These states are also responsible of the well-defined structure seen in Fig. 2(e) between the principal $\bar{\Gamma}$ points. Another sine-like band found experimentally between -0.5 and -0.8 eV (blue dashed line in Fig. 3(c,d)) is identified in the calculations at slightly deeper energies (dashed line in Fig. 3(e)). This state follows the dispersive part of the surface resonance V3/V3' in Fig. 4(e), where two boxes indicate its location along the $\bar{\Gamma}$-$\bar{K}'$ and $\bar{\Gamma}$-$\bar{K}$ directions.

Good agreement between theory and experiment is found also along other symmetry directions of the system. Fig. 5 reports ARPES intensity maps along different directions and for different surface-sensitive photon energies. These data can be compared with the $\rho(E,k_{//})$ of Fig. 4(e). Along $\bar{\Gamma}$-$\bar{K}$ the HH/LH Ge film band reproduces according to the ×4 periodicity of the SBZ, as indicated by the topmost dotted line in Fig. 5(a). The band with maxima at -0.38 eV at the $\bar{\Gamma}$ points is associated with the V2 surface resonance. The intense feature seen at -0.8 eV is identified with the V4 surface resonance, which has a strong component from around the $\bar{K}$ point. Along $\bar{J}$-$\bar{K}$ (Fig. 5(b)) two features displaying a ×5 periodicity are visible. The topmost band is the V1 surface state, while the lower band can be associated to the flat part of the V3 state. Along $\bar{J}$-$\bar{K}'$ two nearly flat bands appear at -0.22 and -0.7 eV (Fig. 5(c-f)). The top one is the V1 surface state, while the bottom one is the V3 surface resonance. The experimental identification of the V1-V4 states, producing distinct peaks in the $\rho(E)$ of Fig. 4(d), demonstrates the validity of the RS structural model [11].

Finally, we comment on the possible origin of the elongated patterns observed in Fig. 3(b) along several $\bar{J}$-$\bar{K}'$ directions. Fig. 5(c,f) demonstrates that along equivalent $\bar{J}$-$\bar{K}'$ lines (i.e., along $[0\bar{1}0]$, the direction parallel to the steps) and for different photon energies the V1 and V3 states are nearly flat. In contrast, Figs. 3(c,d) and 5(b) show that the V1 and V3 states exhibit a distinct dispersive character along the $\bar{\Gamma}$-$\bar{K}'$ and $\bar{J}$-$\bar{K}$ directions, i.e. along $[\bar{5}01]$, the direction perpendicular to the steps. This anisotropic in-plane dispersion cannot arise from a simple band folding: the $\bar{J}$-$\bar{K}'$ reciprocal distance is larger than $\bar{J}$-$\bar{K}$, therefore the folding of the bands would be denser along $\bar{J}$-$\bar{K}$. We notice that the two directions are located at approximately 45° to the light scattering plane, and hence are nearly equivalent from the point of view of the experimental geometry. The strong directional dependence is somewhat unexpected after the considerable reconstruction and rebonding of the (105) steps, which leaves little clear evidence of a structural anisotropy between the [0$\bar{1}$0] and [$\bar{5}$01] directions (see Fig. 1(f,g)), especially if a more realistic diffuse Ge/Si interface is considered [11]. Nonetheless, strong anisotropies are also observed in STM measurements (see Fig. 1(d,e)) and optical spectroscopy [11]. These anisotropies can be traced back to electronic states of the Ge horseshoe motifs. Due to the arrangement of pairs of horseshoes in the cell, there is effectively a better coupling along the [$\bar{5}$01] direction between orbitals associated with the horseshoe "legs". The resulting surface state (V1) is extended along [$\bar{5}$01], yielding a dispersive band (Figs. 3(c,d) and 5(b)) in the ARPES that contributes strongly to the zigzag chain seen in the STM. The lack of coupling between horseshoes along the steps leads to a relatively localized/flat band in the [0$\bar{1}$0] direction. A similar mechanism should explain the anisotropic appearance of V3.

In summary, we used ARPES to characterize in a direct way the electronic band structure of the RS-reconstructed Ge/Si(105) surface. Within 1 eV below the VBM we observed many states displaying the expected 5×4 periodicity and the dispersion predicted by first-principles calculations. The good matching between experiment and theory confirms the validity of the structural model built to explain the optical transitions observed in the RAS spectrum. The elongated photoemission patterns of the V1 and V3 states along the [0$\bar{1}$0] axis are linked to the in-plane anisotropic coupling between orbitals associated with surface horseshoe motif.

**Acknowledgments**

C.H. acknowledges high-performance supercomputing resources and support from CINECA through the ISCRA initiative. L.P. acknowledge the European Union's Horizon 2020 research and innovation programme under grant agreement No. 766719- FLASH project.

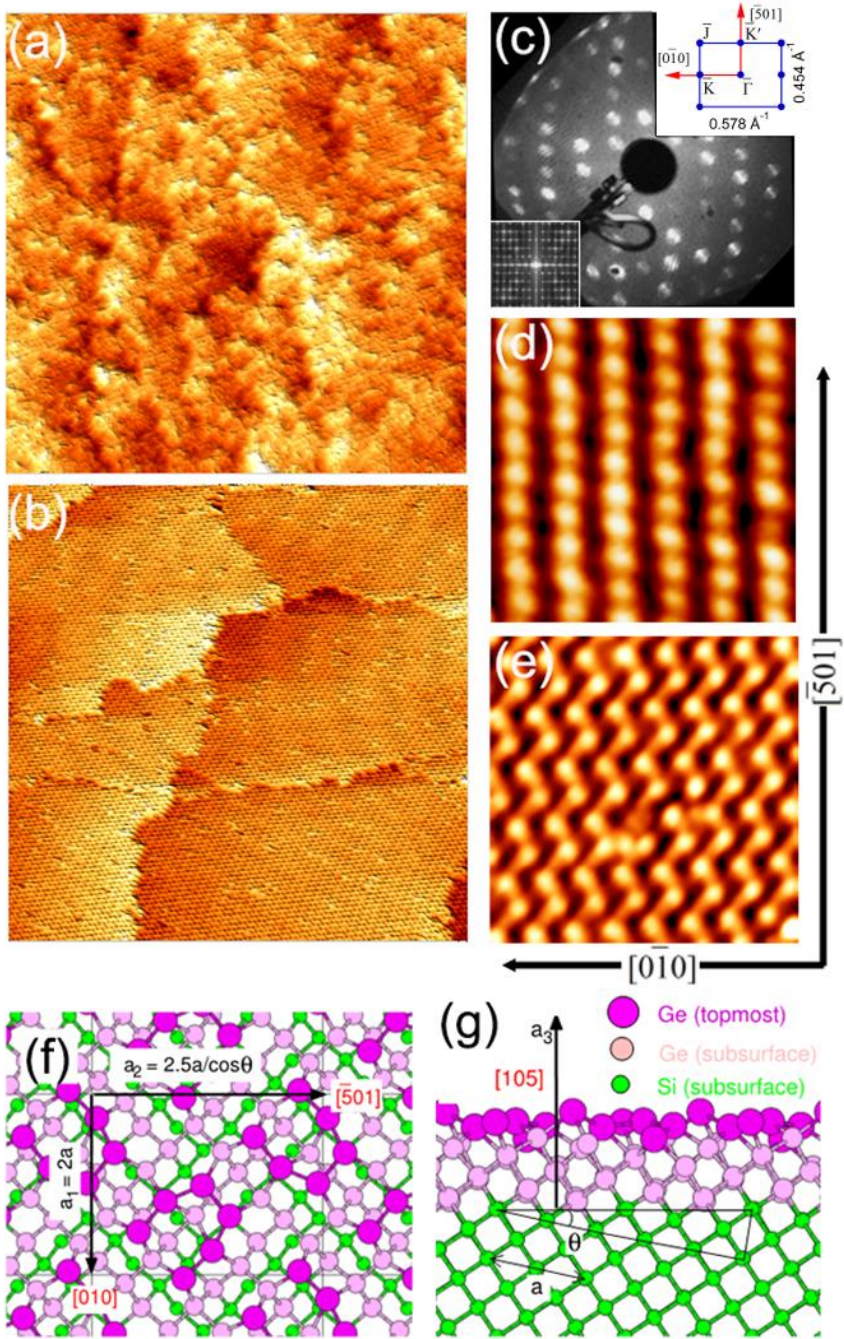

**Figure 1**. (a,b) STM images (88 x 88 nm$^2$) of (a) clean Si(105) (b) 4 ML of Ge/Si(105). (c) LEED pattern of the RS-reconstructed Ge/Si(105) surface (60 eV). In the insets, the SBZ and the simulated LEED pattern are shown. (d,e) Atomic-resolution STM images for (d) empty (sample bias $U$= +2.5V) and (e) filled states (sample bias $U$= -1.5 V) at a constant current of 1.5 nA. (f, g) Structural model of the RS-reconstructed Ge(4 ML)/Si(105) surface: (f) top and (g) side view.

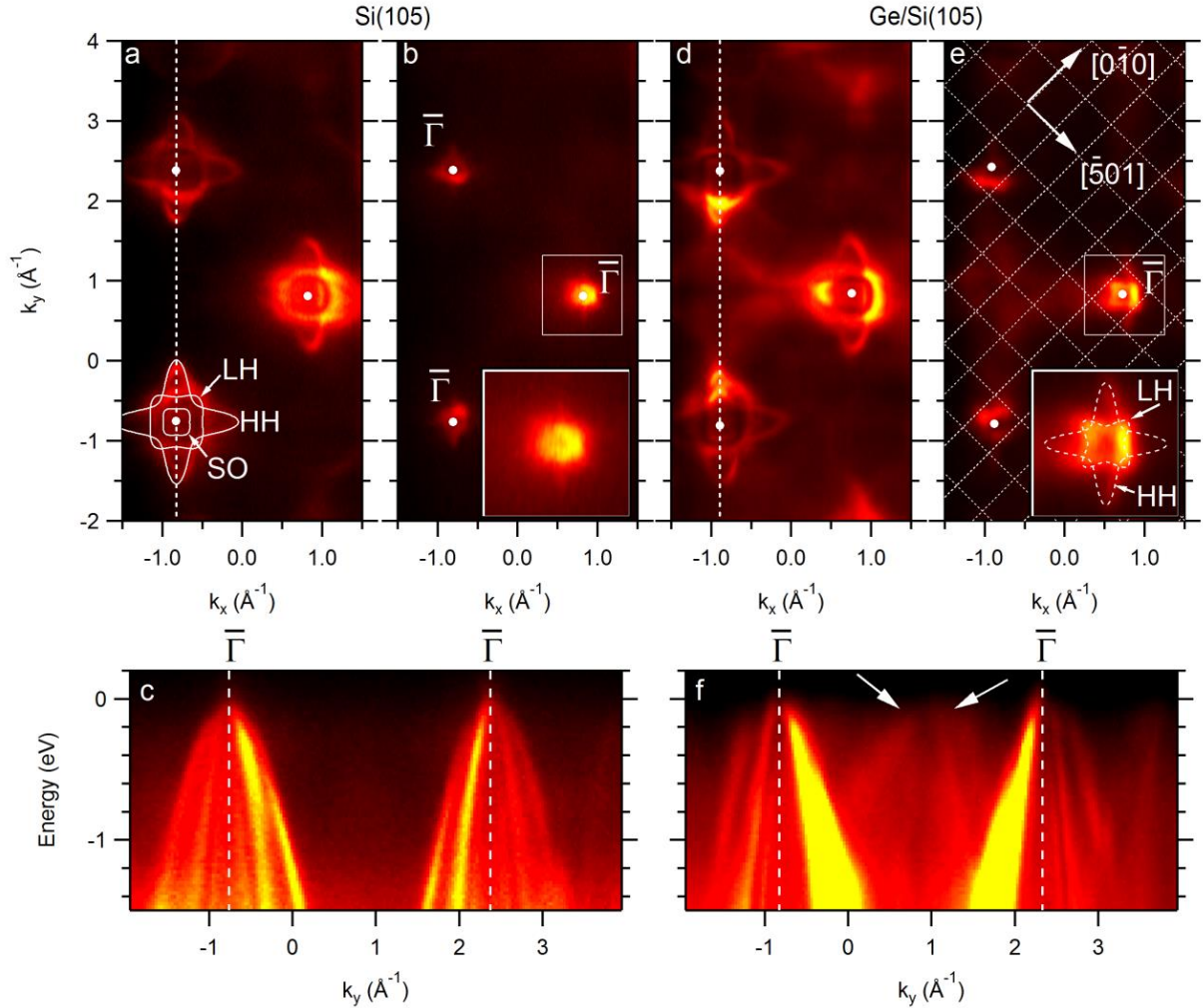

**Figure 2.** ARPES analysis with photon energy 130 eV. (a,b) ARPES constant energy cuts of clean Si(105) taken at (a) -1.00 eV (b) -0.13 eV with respect to the VBM. White continuous lines indicate HH, LH and SO contours. White dots mark the position of $\bar{\Gamma}$ points. The inset shows a zoom near one $\bar{\Gamma}$ point. (c) ARPES data along the direction indicated in panel (a) by a white dashed line. (d,e) ARPES constant energy cuts of 4 ML Ge on Si(105) taken at (d) -1.00 eV (e) -0.13 eV with respect to the VBM. The white dashed grid marks the edges of the repeated 5×4 SBZ. The inset shows a zoom near a $\bar{\Gamma}$ point, in analogy to panel (b). White dashed lines indicate the Ge-related HH and LH contours. (f) ARPES data along the direction indicated in panel (d) by white dashed line. White arrows indicate the appearance of new states, not seen in clean Si.

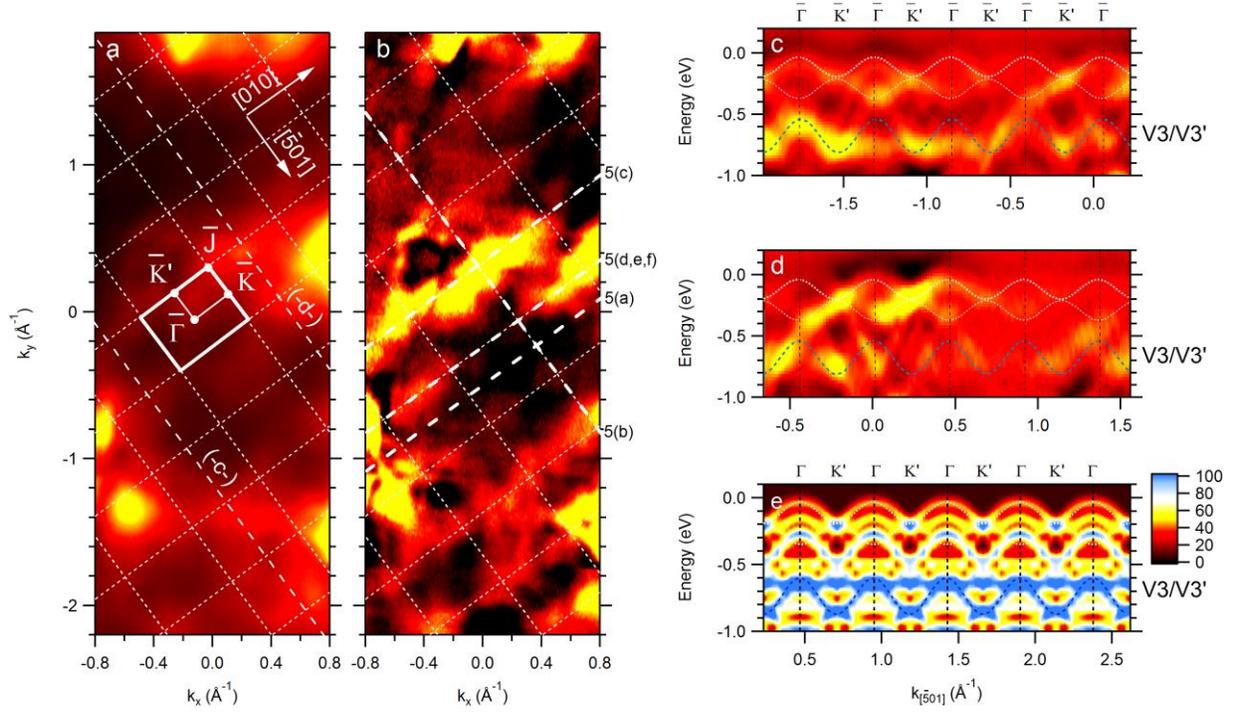

**Figure 3**. (a-d) ARPES analysis of 4 ML Ge on Si(105) with photon energy 40 eV. (a) ARPES constant energy cut at -0.25 eV. The thick rectangle and thin dotted lines indicate the 5×4 SBZs. Dashed lines mark the direction of the cuts shown in panels (c,d). (b) Second derivative along the energy axis of the data shown in panel (a). Bold dashed lines mark the positions of the cuts shown in Fig. 5(a-f). (c,d) Second derivative ARPES data along the lines indicated in (a), and shown in the second derivative form along the energy axis. (e) Calculated band structure along $\bar{\Gamma}$-$\bar{K}'$ direction of the model 4 ML Ge on Si(105). Light blue, yellow, and red colors indicate surface Ge, mixed Ge/Si, and bulk Si states, respectively.

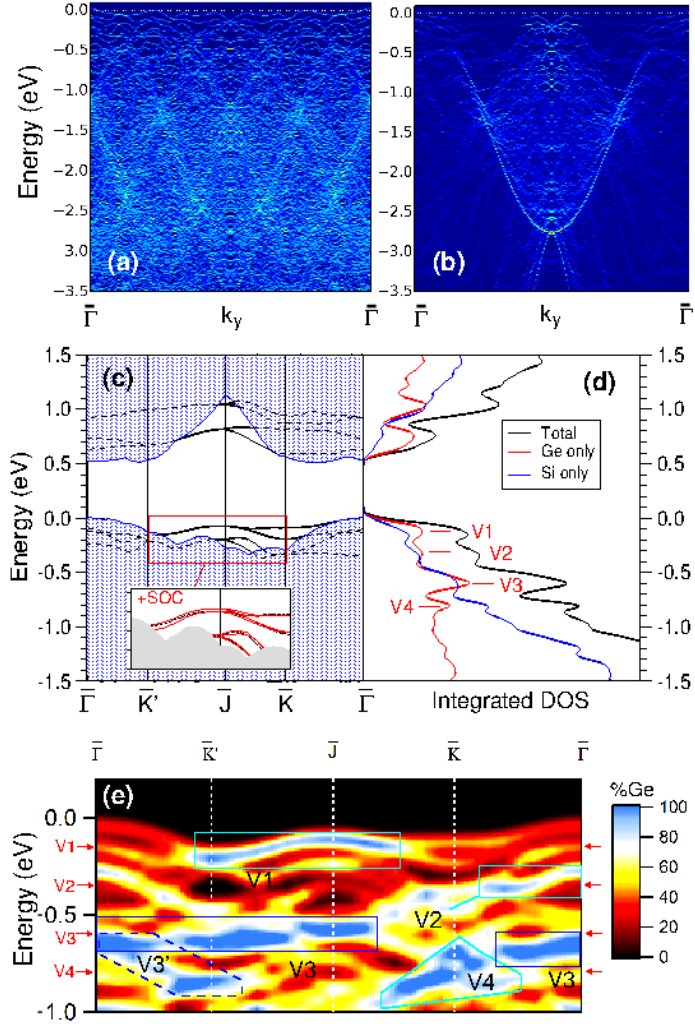

**Figure 4**. Calculated band structure of the model 4 ML Ge on Si(105). (a,b) Band dispersion along the $k_y$ direction scanned in Fig. 2(f) (a) within the repeated zone scheme and (b) following an unfolding procedure (see text). (c) Band structure around a closed path in the SBZ. Shaded areas indicate the projected bulk bands of silicon. Inset: red and black lines indicate surface states calculated with and without including spin-orbit coupling. (d) Integrated DOS projected on Ge and Si atoms and their sum (Total). Four main peaks in the valence band DOS are indicated as V1-V4. (e) $k$-resolved DOS $\rho(E,k_{//})$ around the same path as (c). The color scale indicates the weight on Ge atoms and allows to identify the location of the V1-V4 states in k-space (arrows indicate the peak energies as found in the integrated DOS (d)).

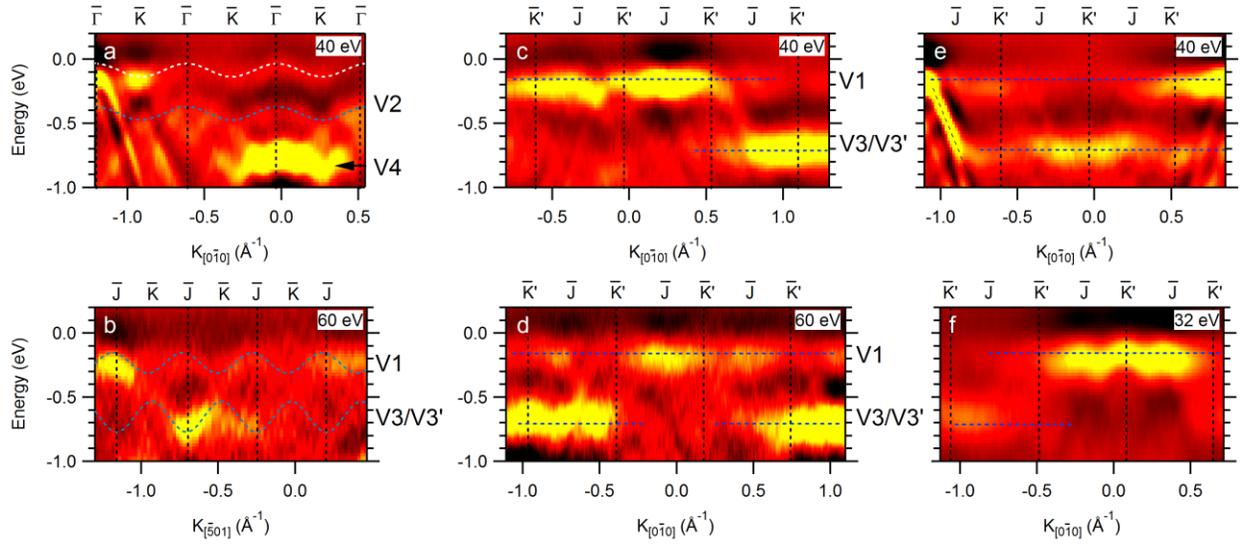

**Figure 5**. Second derivative ARPES data taken along the lines marked in Fig. 3(b) with photon energies indicated in the panels. (a) $\bar{\Gamma}$-$\bar{K}$, 40 eV (b) $\bar{J}$-$\bar{K}$, 60 eV (c) $\bar{J}$-$\bar{K}'$, 40 eV (d) $\bar{J}$-$\bar{K}'$, 60 eV (e) $\bar{J}$-$\bar{K}'$, 40 eV (f) $\bar{J}$-$\bar{K}'$, 32 eV.